\documentclass[twocolumn]{aa}
\usepackage[dvips]{graphicx}
\usepackage[english]{babel}
\usepackage[latin1]{inputenc}
\usepackage{latexsym}
\usepackage{rotating}
\usepackage{amssymb,amsmath}

\usepackage{natbib}
\def\st{\scriptstyle}

\def\p{\ifmmode\pm\else$\pm$\fi}

\def\msun{\ifmmode \rm{M}_\odot \else M$_\odot$\fi} 
\def\mdot{\ifmmode \dot M \else $\dot M$\fi}    

\newcounter{sub}
\newcounter{subeqn}[sub]
\renewcommand{\thesubeqn}{\alph{subeqn}}
\renewcommand{\theequation}{\thesub\thesubeqn}

\def\lp{\left(}
\def\rp{\right)}

\def\st{\stepcounter{sub}}
\def\stq{\stepcounter{subeqn}}
\def\be{\begin{equation}}
\def\ee{\end{equation}}
\def\bea{\begin{eqnarray}}
\def\eea{\end{eqnarray}}
\def\bean{\begin{eqnarray*}}
\def\eean{\end{eqnarray*}}
\newcommand\xxi{{\mbox{\boldmath $\xi$}}}
\newcommand\OOmega{\mbox{\boldmath $\Omega$}}
\newcommand\pphi{\mbox{\boldmath $\phi$}}
\newcommand\nab{\mbox{\boldmath $\nabla$}}

\def\v{{\bf v}}
\def\r{{\bf r}}

\newcommand\B{{\bf B}}

\newcommand\no{\nonumber}

\begin{document}

\title{Quasi-periodic oscillations:\\
Resonant shear Alfv\'en waves in neutron star magnetospheres}
\author{V. Rezania$^{1,2}$ \and J. C. Samson$^1$}
\offprints{V. Rezania\\ email: vrezania@phys.ualberta.ca\\
                  $^*$ Present address}
\institute{ $^1$Theoretical Physics Institute,
    Department of Physics,
    University of Alberta\\
    Edmonton, AB, Canada, T6G 2J1$^*$\\
$^2$Institute for Advanced Studies in Basic Sciences,
          Zanjan 45195, Iran}
\date{Received / accepted }
\titlerunning{Resonant Alfv\'en waves in neutron star magnetospheres}
\authorrunning{Rezania \& Samson}

\abstract{
 In this paper we propose a new model for quasi-periodic oscillations
(QPOs) based on oscillating magnetohydrodynamic modes in neutron
star magnetospheres. We argue that
the interaction of the accretion disk with the magnetosphere can
excite resonant shear Alfv\'en waves in a region of enhanced
density gradients, the region where accretion material flows along the
magnetic field lines in the magnetosphere (see Fig. \ref{star-fig1}).
We demonstrate that depending on the distance
of this region from the star and the magnetic
field strength, the frequency of the field line resonance can
range from several Hz (weaker field, farther from star), to
approximately kHz frequencies (stronger field, $\sim 2-10$ stellar
radii from the star).  We show that such oscillations are able to
significantly modulate inflow of matter from the high density
region toward the star's surface, and possibly produce the observed
X-ray spectrum. In addition, we show that the observed $2:3$
frequency ratio of QPOs is a natural result of our model.

}

\maketitle
%
\section{Introduction}

The discovery of quasi-periodic oscillations (QPOs) in low-mass
X-ray binaries (LMXBs) has been reported and discussed in recent
studies in theoretical and observational astrophysics. The {\it
Rossi X-Ray Timing Explorer} has observed oscillations in the
X-ray flux of about 20 accreting neutron stars. These oscillations
are very strong and remarkably coherent with frequencies ranging
from $\sim 10$ Hz to $\sim 1200$ Hz.  The kHz frequencies
correspond to dynamical
time scales at radii of a few tens of kilometers, and are possibly
closely related to the Keplerian orbital frequency of matter at
the inner disk. Almost all sources have also shown twin spectral
peak QPOs in the $500$ Hz to $1200$ Hz part of the X-ray spectrum,
and both peaks
are moving up and down in frequency simultaneously.

Although there are some detailed differences in different types of
X-ray sources, the observed QPOs
are remarkably similar, both in frequency and peak separation.
(Six of the 20
known sources are originally identified as ``Z'' sources and the
rest are known as ``atoll'' sources (see table 1). For more
information about the atoll and Z sources, see van der Klis 2000).
Such a similarity shows that QPOs should depend on general
characteristics of the X-ray sources which are common to all
systems.  In other words, the QPO can be regarded as a generic
feature of the accreting neutron star.

Remarkably, the characteristic
dynamical time scale for material orbiting near the compact object
is comparable to the observed millisecond X-ray variability, ie.
because $\tau_{\rm dyn}=\sqrt{r^3/(G M)}\sim 2~ {\rm ms}~ (r/100 {\rm
km})^{3/2} (1.4 M_\odot/M)^{1/2}$ where $r$ is the distance of
the orbiting material from the center of the compact object with mass $M$ (circular orbit),
$G=6.67\times 10^{-8}$ dyn cm$^2$/g$^2$, and $M_\odot=1.99\times 10^{33}$ g.
Such a natural time scale is
the foundation of most models for the observed QPOs.
\cite{AS85} first proposed a beat-frequency model to explain the
low-frequency\footnote{Two different low-frequency ($<100$\,Hz)
QPOs were known in the Z sources, the $6-20$\,Hz so-called normal
and flaring-branch oscillation (NBO; \citealt*{MP86}) and the
$15-60$\,Hz so-called horizontal branch oscillation (HBO;
\citealt*{Van85}).} horizontal branch oscillation (HBO) seen in Z
sources (see also \citealt*{Lam85}).  They used the Alfv\'en radius,
where the magnetic pressure balances the ram pressure of
the infalling material in spherical accretion,
$r_{\rm A} \sim 1.5 \times 10^6~{\rm cm}~\dot{M}_{17}^{-2/7}
\mu_{26}^{4/7} (M/M_\odot)^{-1/7}$, as the preferred
radius.   For this Alfv\'en radius, the assumptions are:
magnetic field is stellar dipole, velocity and mass density are for free-fall.
Here $\dot{M}_{17}$ is the mass accretion rate (this corresponds to
$1.5\times 10^{-9} M_\odot$yr$^{-1}$ and is typical for low-mass accreting neutron
stars)
in units of $10^{17}$ g s$^{-1}$ and $\mu_{26}$ is the magnetic
dipole moment at the stellar surface in units of $10^{26}$ G cm$^3$.
Note that we assumed $\dot{M}$ is constant in time and position.

\cite{Mil96,Mil98a} proposed a beat-frequency model to study
kHz QPOs, the so-called
the sonic-point beat-frequency model, based on a new preferred
radius, the sonic radius.  In this model, they assumed that near
the neutron star there is a very narrow region of the disk in
which the radial inflow velocity increases rapidly as the radius
decreases. Such a sharp transition in the radial velocity of
plasma flow from subsonic to supersonic happens at the ``sonic
point'' radius, $r_{\rm sonic}$.  This radius tends to be close to
the innermost stable circular orbit (ISCO) \footnote{In general
relativity, no stable orbital motion is possible within the
innermost stable circular orbit (ISCO), $r_{\rm ISCO} = 6GM/c^2
\approx 12.5 M_{1.4M_\odot}\,\hbox{km}\,$. The frequency of
orbital motion at the ISCO, the highest possible stable orbital
frequency, is $\nu_{\rm ISCO} \approx (1580/M_{1.4M_\odot})\,
\hbox{Hz}$.}, $r_{\rm ISCO}$, however, radiative stresses may
change its location, as required by the observation that the kHz
QPO frequencies vary with time.  Comparing the HBO and kHz QPO frequencies,
clearly $r_{\rm sonic}\ll r_{\rm A}$, so part of the accreting
matter must remain in near-Keplerian orbits well within $r_{\rm
A}$.
An important prediction of the sonic-point model is that
$\Delta\nu = \nu_2-\nu_1$ be constant at the spin
frequency of the neutron star $\nu_s$, which is
contrary to observations \citep{LM99}.   Accurate observations from
the millisecond pulsar SAX J1808.4-3658 show that the frequency separation
$\Delta\nu=195$ Hz,
while the pulsar spin frequency is $\nu_s=401$ Hz \citep{Wij03,Cha03}.
Nonetheless, further observations reveal that the frequency
separation $\Delta\nu$ decreases when $\nu_2$ increases.

\cite{SV98,SV99} proposed a relativistic
precession model in which the high kHz QPO frequency $\nu_2$ is
identified with the Keplerian frequency of an orbit in the disk
(similar to the beat-frequency model) and the low kHz QPO
frequency $\nu_1$ with the periastron precession of that orbit.
Furthermore, in the relativistic model $\Delta\nu$ and $\nu_{s}$
are not expected to be
equal as in a beat-frequency interpretations. An interesting
result of the relativistic precession model is that $\Delta\nu$
should decrease not only when $\nu_2$ increases (as observed) but
also when it sufficiently decreases.
However, in this model it is not clear how the X-ray flux is modulated
at the these frequencies.

Recently \cite{Zha04} proposed a model for kHz QPOs based on
magnetohydrodynamic (MHD)
Alfv\'en oscillations in the disk.  He introduced a new preferred radius, the
quasi-sonic point radius, where the Alfv\'en velocity matches
the orbital Keplerian velocity. The author suggested that the upper and lower
kHz frequencies are the MHD Alfv\'en wave frequencies
at the quasi-sonic point radius with low and high
accreted material mass densities, respectively.
In this model,
the frequency separation will decrease when $\nu_1$ increases
and/or sufficiently decreases.
Although his results show good agreement
with observations, several vital issues are left with no answers.
For example, no mechanism is suggested to explain the
excitation of Alfv\'en wave oscillations within the star's magnetosphere.
More importantly, it is not clear how these Alfv\'en wave frequencies modulate
the X-ray flux coming from the surface of the star.

\begin{figure}[ht]
\centerline{\includegraphics[width=.8\hsize]{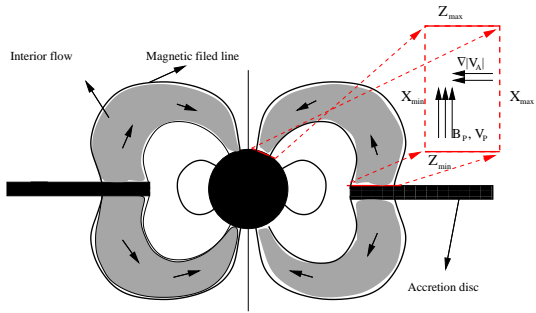}}
\caption{A schematic and idealized side view of an accreting neutron star-disk
system.  The neutron star's strong gravity causes
a very high velocity flow toward the magnetosphere.  As a result,
the magnetosphere is pushed inward in the disk plane but balloons
outward in direction away from the disk plane. Some of the plasma
may leave the disk and flow along the field lines.  The magnetic
star-disk connection by a dipolar
magnetosphere as well as the polar accretion flow along the field lines
are part of a reasonable but simplified picture of magnetized
star-disk systems in general.  The dashed box represents the model that we
considered in this paper.  $z_{max}>0$ represents the surface of the star,
$z_{min}>0$ is on the disk plane, $x_{min}$ and $x_{max}$ are embeded the
density gradiant region.
The directions of the ambient magnetic field and the plasma flows in the
box are
indicated by the verticle arrows, while the direction of the gradient in
the Alfv\'en speed is shown by the horizontal arrows.
}
\label{star-fig1}
\end{figure}

In this paper, we propose a new model to explain QPOs in LMXBs
based on interaction of accreting
plasma with the neutron star magnetosphere.  This interaction is a
common feature in all accreting binary systems.   Matter transfers
from the companion star to the compact
object. This matter is accelerated by the gravitational pull of
the compact object and hits the magnetosphere of the star with a
sonic/supersonic speed. The MHD interaction of the infalling
plasma with the neutron star magnetosphere, will alter not only
the plasma flow toward the surface of the star, as assumed by
current QPO models, but also the structure of the star's
magnetosphere.  The magnetic field of the neutron star is
distorted inward by the infalling plasma of the Keplerian
accretion flow. Since the gravity of the star confines the inward
flow to a small solid angle $\sim 10^{-2}$ ster, the magnetic
field of the star will be more compressed in the disk plane than
in other areas (see Fig. \ref{star-fig1}). Furthermore, in a more
realistic picture, one would expect that the highly accelerated
plasma due to the infall process would be able to penetrate
through the magnetic field lines. Such material will be trapped
by the magnetic field lines and produce enhanced density
gradient regions
within the magnetosphere.
\footnote{Note that the enhanced density
gradient region is defined as follow: when we move outward from the surface
of the star, the plasma density in the magnetosphere is more or
less is constant, say $\rho_0$.
At the point that magnetosphere interacts with the accretion disk
(more likely at the disk's inner edge),
however, there is a sudden increase in the magnetospeheric plasma density,
say $\rho>\rho_0$, due to
the motion of accretion materials along field lines. After this thin region (comparing
with whole magnetosphere), the
magnetospheric plasma density falls down to $\rho_0$ again.  Therefore,
the enhanced density gradient
region is not a region with a gradient in the plasma density, only. It is a region
with different (higher) plasma density compared with the rest of the magnetosphere.}
However, due to the material's initial
velocities, the penetrating material moves along the field lines,
finally hitting the star's
surface at the magnetic poles and producing the observed X-ray
fluxes (Fig. \ref{star-fig1}). See \cite{GLP77} for more detail.

Besides modifying the geometry of the magnetosphere, the compressional
action of the accretion flow can excite some plasma perturbations in the
region of enhanced density gradients. This can be understood by noting the
fact that the inward motion of the accretion flow will be halted
by the outward magnetic pressure at a certain distance from the
star, the Alfv\'en radius \citep{GLP77}. In other words, the
accretion flow pushes the stellar magnetic field toward the star
until the pressure of the infalling plasma $\rho v_r^2/2$
balances the magnetic pressure $B_p^2/8\pi$.  Here
$\rho$ and $v_r$ are the density and radial velocity of the
infalling matter and $B_p$ is the poloidal magnetic field at the
disk plane. Therefore, any instability in the disk at the Alfv\'en
radius would disturb such quasi-equilibrium configurations as well
as the structure of the magnetosphere,
possibly through the Rayleigh-Taylor and/or ballooning
instability.
As an example, the interaction
of the solar wind with the Earth's magnetosphere excites resonant shear
Alfv\'en waves, or field line resonances (FLRs), along the
magnetic field lines \citep{Sam91}.  As a result, one might expect
that such Alfv\'en waves can be excited by the compressive
accreting plasma in the magnetosphere of an accreting neutron star
with the accretion flow playing the role of the solar wind.

The plan of this paper is as follow.  We outline the occurrence of
resonant coupling in a magnetized plasma in section
\ref{sect:mag}.   We review the excitation of shear
Alfv\'en waves by studying linear perturbations in MHD.  Next, the
occurrence of FLRs, as a result of a
resonant coupling between the compressional and shear Alfv\'en
waves, is discussed.  To illustrate the basic features of FLRs,
the excitation of these resonances in a rectilinear magnetic field
is considered.  In section \ref{amb}, in order to consider a
more realistic model for an accreting neutron star, we study the
influence of the ambient flow along the field line on the
excitation of FLRs. We demonstrate that in this case the
eigenfrequency of the Alfv\'en modes is modulated by the velocity
of the field aligned plasma flow.
Depending on the field line where the resonance
occurs, the eigenfrequency of the FLR is in the range of several
hundred Hz to kHz.
Furthermore, in the presence of this flow,
the plasma displacement parallel to the magnetic
field lines is non-zero.  This is very important because these displacements,
which are missing in
the case of zero ambient flow, might be responsible for modulating the
motion of infalling material toward the magnetic poles and
producing the observed X-ray fluxes.  Existence of this non-zero parallel component
distinguishes our study from other related investigations in which the modulation
of X-ray flux in observed QPOs is fairly unclear or ambiguous.
A possible occurrence of more
than one peak in the power spectrum is also discussed. In
addition, we show that the observed $2:3$ frequency ratio of QPOs
is a natural result of our model. Section \ref{conc} is devoted to
summarizing our results and further discussions.

%

\section{The magnetospheric model}\label{sect:mag}


\subsection{Magneto-hydrodynamic waves}

In general, the dynamics of an ideal magnetized plasma is
described by plasma density $\rho$, plasma pressure $p$,
gravitational potential $\Phi$, velocity vector $\v$ and magnetic
field $\B$:
\st\begin{eqnarray}\label{eqm} \stq\label{euler}
&&\partial\v/\partial t + \v \cdot \nab \v =
 - \nab p/\rho  - \frac{1}{4\pi\rho}\B \times (\nab \times \B) - \nab\Phi,\no\\
 &&\\
\stq\label{continuty}
&&\partial\rho/\partial t + \nab\cdot(\rho \v)=0,\\
\stq
&&\partial \B/\partial t = \nab\times (\v \times \B), \label{mhd_ind_1} \\
\stq
&&\nab\cdot \B = 0\,. \label{div_B}
\end{eqnarray}
An adiabatic equation of state with an adiabatic index $\gamma$,
$p/\rho^\gamma=$const., is assumed in this paper in order to
complete the above equations.

In a steady state (ie. $\partial/\partial t=0$), Eqs. (\ref{eqm})
have been studied in detail in connection with the problem of stellar
winds from rotating magnetic stars \citep{Mes68} and in
connection with diffusing/flowing plasma into magnetospheres in
accreting neutron stars \citep{EL77,EL84,GLP77}. Obviously, in
the case of accreting neutron stars, $\v$ represents the inflow
velocity of matter accreted to stars, while in stars with a
stellar wind it represents the plasma outflow velocity.

Consider an axisymmetric system consisting of
a rotating star with a constant angular velocity
$\Omega_s$ along the z-axis and a magnetic field
whose symmetry axis is aligned with
the rotation axis of the star.
By decomposing the velocity and magnetic field vectors into poloidal
and toroidal components:
\st \be\label{comp} \v=\v_p + \Omega
\varpi \hat{\pphi},~~~~~~~\B=\B_p + B_\phi\hat{\pphi},
 \ee
one can obtain
\bea \st\label{vp}
&&\v_p=(f/\rho)\B_p,\\
\st\label{om}
&&\Omega=\Omega_s+(f/\rho)(B_\phi/\varpi),
\eea
where $f$ is the mass flux along a magnetic flux tube of unit flux
\citep{GLP77}.
The subscripts $p$ and $\phi$ denote
poloidal and toroidal components, respectively, $\Omega$ is the
angular velocity of the plasma, $\varpi$ is the
distance from the axis of the rotation, and
$\hat{\pphi}$ is a unit toroidal vector.   We note that both $f$ and
$\Omega_s$ are constant along a given field line. Equation
(\ref{om}) can be rewritten as
\st \be
\Omega=\Omega_s+(v_p/\varpi)(B_\phi/B_p),
 \ee
where $v_p=\v_p\cdot\hat{\bf p}$ is the magnitude of velocity along the
poloidal magnetic field $\B_p$ with magnitude $B_p$ and $\hat{\bf p}=\B_p/B_p$.
Furthermore, \cite{GLP77} have shown that at distances close to
the Alfv\'en radius, the magnitude of the poloidal component of
the inflow velocity (for accreting systems) $v_p$ approaches the
poloidal Alfv\'en speed $v_{\rm A}$, i.e. \footnote{ In order to estimate
the poloidal component of the inflow velocity $v_p$, one needs to
integrate the momentum Eq. (\ref{euler}):
\st \be\label{vp_1}
(1/2)(v_p^2 + \Omega^2\varpi^2)-\Omega_s\Omega \varpi^2 -GM/r={\rm
 constant~ along~ a~ given~ field~ line},
\ee
where $M$ is the mass of the neutron star
\citep{Mes68,GLP77}. In the above equation the pressure term is
neglected.  Equation (\ref{vp_1}) shows conservation of energy in
a frame corotating with the star, while in a nonrotating frame the
extra term $\Omega_s\Omega\varpi^2$ appears, that represents the
work done by the magnetic field on the flowing plasma. However, as
argued by \cite{GLP77} the poloidal velocity in the inner magnetosphere
where $r\ll r_{\rm A}$ is nearly equal to the free-fall velocity, i.e.
\st
\be\label{vp1} v_p\sim (2GM/r)^{1/2}. \ee }
\st \be v_p^2(r_{\rm
A})=v_{\rm A}^2(r_{\rm A}).
\ee

The possible perturbations of a magnetized plasma and MHD waves
are found by specifying the equilibrium configuration of the star
and then solving the perturbed Eqs. (\ref{eqm}). This is a nontrivial problem,
that is addressed by several investigators. However, in this paper
we will be interested in the propagation of shear Alfv\'en waves in
the star's magnetosphere and the resulting FLRs.


\subsection{Field line resonances}

In MHD, waves can propagate in three different modes including
shear Alfv\'en waves, and the fast and slow compressional modes
\citep{LL92}. In a homogeneous plasma, one can easily show that
these three modes are linearly independent. However, in an
inhomogeneous medium these three modes can be coupled, yielding
either a resonant coupling \citep{Sou74,Has76}, or an unstable
ballooning mode \citep{OT93,Liu97}. FLRs result from the coupling
of the fast compressional and the shear Alfv\'en modes
\footnote{In the cold-plasma approximation, the fast compressional wave
(fast magnetosonic wave) is also called the compressional Alfv\'en
wave.}
whereas the ballooning instability results from the coupling
of the slow compressional and the shear Alfv\'en modes. However,
in the cold-plasma approximation (which is appropriate to
neutron star magnetospheres), the slow compressional wave does not exist.
FLRs are standing waves that are stimulated
in a pulsar with gradients in the Alfv\'en speed transverse
to the ambient B-field, particularly, within density boundary
layers (a parallel gradient in the Alfv\'en speed is also included).
Efficient coupling between the shear Alfv\'en wave and
the fast compressional
wave can produce a relatively narrow FLR spectrum,
even when the driver is broadband.

The linear theory of the FLRs was developed by \cite{CH74} and
\cite{Sou74}, and applied to auroral phenomena by \cite{Has76}.
\cite{Sam03} developed a nonlinear model with a nonlocal
electron conductivity to explain the evolution of field aligned
potential drops and electron acceleration to form  auroral arcs. They
studied the possible coupling between the fast compressional mode
and the shear Alfv\'en mode in an inhomogeneous plasma with a radial
gradient in the Alfv\'en velocity $v_{\rm A}=B/\sqrt{4\pi \rho}$.
For further discussion of mode conversion for shear Alfv\'en waves see for
instance \cite{Sti92}, and for an example of computational models
see \cite{RW94}.  Analytic solutions for toroidal FLRs in dipole magnetic fields
can be found in \cite{TW84}.

To illustrate the resonant coupling between the fast compressional
mode and the shear Alfv\'en mode we study the excitation of FLRs
in an inhomogeneous plasma in a rectilinear magnetic field model.


\subsubsection{FLRs in a rectilinear magnetic field}
Plasma dynamics can often be described by the adiabatic ideal MHD equations:
\st
\begin{eqnarray}\label{mhd_eq}
\stq
&&\rho( \partial/\partial t +\v\cdot\nab)\v =
- \nab p - {1\over 4\pi} \B \times (\nab\times \B), \label{mhd_mom} \\
\stq
&&\partial \B/\partial t = \nab\times (\v \times \B), \label{mhd_ind} \\
\stq
&&\nab\cdot\B = 0, \label{mhd_div} \\
\stq
&&\partial \rho/\partial t + \nab \cdot (\rho \v) = 0, \label{mhd_cont} \\
\stq
&&d(p\rho^{-\gamma})/dt=0. \label{mhd-stat}
\end{eqnarray}
where we neglect the effect of gravitational attraction on the
plasma (see Eqs. (\ref{eqm}) for comparison). Introducing
Eulerian perturbations to the ambient quantities in the form
\begin{eqnarray}
&& p = p_0 + \delta p,~~ \rho = \rho_0 +\delta \rho, \no\\
&&\B = \B_0 + \delta \B,~~ \v=\delta\v = \partial\xxi/\partial t,\no
\end{eqnarray}
we can derive a linear wave equation by simplifying Eqs.
(\ref{mhd_eq}) up to the first order in the perturbations.
We choose coordinates such that the ambient magnetic field is in the
z-direction.  Without loss of generality, we assume that the
gradients in all the unperturbed quantities are in the x-direction. Setting
the plasma displacement as \st
\begin{eqnarray}
\stq
&&\xxi(\r,t) = \xxi(x) e^{-i (\omega t - k_y y - k_z z)},\\
\stq
&&\delta\B(\r,t) = \delta\B(x) e^{-i (\omega t - k_y y - k_z z)},\\
\stq
&&\delta \rho(\r,t) = \delta\rho(x) e^{-i (\omega t - k_y y - k_z z)},\\
\stq
&&\delta p(\r,t) = \delta p(x) e^{-i (\omega t - k_y y - k_z z)},
\end{eqnarray}
Eqs. (\ref{mhd_eq}) reduce to \citep{HS92}
\st\begin{equation} \label{har}
\frac{d^2 \xi_x}{dx^2} + \frac{F'(x)}{F(x)} \frac{d \xi_x}{dx} + G(x) \xi_x = 0,
\end{equation}
with
\st\begin{equation}\label{G}
G(x) = \frac{\omega^2 \Bigg(\omega^2 - (v_{\rm S}^2+v_{\rm A}^2)(k_y^2 + k_z^2)\Bigg) +
k_z^2 (k_y^2 + k_z^2) v_{\rm A}^2 v_{\rm S}^2}{(v_{\rm S}^2+v_{\rm A}^2)\omega^2 -
k_z^2 v_{\rm A}^2 v_{\rm S}^2},
\end{equation}
and
\st\begin{equation}\label{F}
F(x) = \frac{\rho_0(\omega^2 - k_z^2 v_{\rm A}^2)}{G(x)},
\end{equation}
where $v_{\rm A}^2(x) = B_0^2/(4\pi\rho_0)$ and $v_{\rm S}^2(x)
= \gamma p_0/\rho_0$, and $'=d/dx$.  Here we assumed that the total pressure
(fluid+field) $P_0=p_0+B_0^2/(8\pi)$ is constant.
Equation (\ref{har}) yields two turning
points at $G(x) = 0$,
and two resonances at $F(x) = 0$.
Close to the resonance positions, Eq. (\ref{har}) can be
approximated by \st
\begin{equation}
\frac{d^2 \xi_x}{dx^2} + \frac{1}{x - x_0} \frac{d \xi_x}{dx} + G(x) \xi_x = 0,
\end{equation}
where $x_0$ is the position where the resonance occurs.

In the case of a strong magnetic field, or for a cold plasma (ie.
$p_{\rm fluid}/p_{\rm magnetic}\ll 1$), one can put $v_{\rm S}
\simeq 0$. Then $G(x)$ and $F(x)$, Eqs. (\ref{G}) and (\ref{F}), reduce
to \st
\begin{equation}\label{G1}
G(x) = \frac{\omega^2}{v_{\rm A}^2} - k_y^2 - k_z^2
\end{equation}
and
\st
\begin{equation}\label{F1}
F(x) = \frac{v_{\rm A}^2 \rho_0(\omega^2 - k_z^2 v_{\rm A}^2)}{\omega^2 -
(k_y^2 + k_z^2) v_{\rm A}^2}.
\end{equation}
In this case Eq. (\ref{har}) has only one resonance (the Alfv\'en resonance) at
\st
\begin{equation}\label{flres}
\omega^2 - k_z^2 v_{\rm A}^2=0,
\end{equation}
which corresponds to the dispersion relation for the shear
Alfv\'en wave along the field line.
It is necessary to note that in above calculations four boundary
conditions are assumed.  In the x-direction the boundaries, say,
at $x_{min}=0$ and $x_{max}=x_m$ are considered perfectly reflecting
(i.e. $\xi_x=0$ at boundaries).
Further, we assume that neutron star and accretion disk surfaces
are perfectly conducting so that the
displacement $\xxi$ vanishes at these surfaces.  In the rectilinear model
we represented these boundaries by planes at constant $z$, i.e.
$z_{min}>0$ on the disk and $z_{max}>0$ on the star's surface (see Fig. \ref{star-fig1}).

Briefly, the FLR mechanism can
be outlined as follows: the incoming
compressional wave hits the stellar magnetosphere and reaches
a field line (or a magnetospheric shell) in which the frequency of the
incoming wave matches the eigenfrequency of the standing shear
Alfv\'en wave along that particular filed line.
The resulting forced-oscillating system causes the amplitude of
the shear Alfv\'en wave to grow in time forming FLRs.

The FLR mechanism is generic and likely to occur in many
astrophysical magnetospheres. As a result, one would expect that
FLRs likely occur not only in the Earth's magnetosphere but
also in the magnetospheres of accreting neutron stars. In the case
of the Earth's magnetosphere, the source of energy for the
FLRs is the interaction of the solar wind with the
magnetosphere \citep{HS92}. In accreting neutron stars, the
accreted plasma interacts with the stars' magnetosphere, allowing
the compressional mode to propagate into the magnetosphere and
flow along the field lines toward the magnetic poles. Such a
compressional action of the accretion flow can excite resonant
shear Alfv\'en waves in the enhanced density regions filled by
plasma flowing along the field lines. In section \ref{amb} we
consider this mechanism more carefully to address its possible
relation to the QPOs observed in
accreting neutron stars in LMXBs.
%
%
%
\section{FLRs in the presence of ambient flow
(along the magnetospheric field lines)}\label{amb}
As discussed above, FLRs have been used to model electron acceleration
and auroral arcs in the Earth's magnetosphere. Although
FLRs are likely to be excited in any
magnetospheric system with an input of compressional energy, one must
carefully evaluate the differences between the Earth's and neutron
star magnetospheres.  In neutron star magnetosphere, for example,
the strong magnetic field
of the neutron star, the rapid stellar rotation, and the intense radiation
pressure from the stellar surface should be considered.

In the case of the Earth's magnetosphere,
due to the small gravitational attraction of the Earth and also
its large distance from the Sun, the solar wind impacts
the whole Sunward side of the geomagnetosphere.
This supersonic solar wind produces the
so-called bow shock structure at the outer boundary of the geomagnetospher.
On the contrary, the
strong gravity of the neutron star creates a converging flow that is supersonic
long before the flow hits the star's magnetosphere. Such a
localized flow is able to change the structure of the
magnetosphere in local areas, particularly in the equatorial plane.
In addition, the highly variable nature of the exterior flow can
change the magnetosphere's structure dramatically in time.
Large flux of plasma stresses the star's outer magnetosphere and
creates a relatively high plasma density in this region.
The plasma then flows along the field lines, an interior
flow, until it hits the star's surface near the magnetic poles
(see Fig. \ref{star-fig1}).
In this section we study the
excitation of FLRs by considering such plasma flows in the
neutron star magnetosphere.
In the context of the Earth's magnetosphere, however,
\cite{HS92} studied the resonant excitation of
Alfv\'en waves by surface waves excited by the solar wind and plasma
flows in the Earth's magnetosheath.

The presence of a flow $\v$ in the plasma adds more modes to the
plasma waves. In general, such flows are a combination of plasma
flow along the magnetic field lines, $\v_p$, and rotational motion
of the plasma around the star, with angular velocity $\OOmega$, ie.
$\v=\v_p+\OOmega\times\r$ (see Eq. \ref{comp}).
As we shall see in Eq. (\ref{par-comp}), the existence of
$\v_p$ alone induces a non-zero parallel component (relative to
the direction of the magnetic field) of displacement
that vanishes in a cold plasma
if $\v_p=0$. We note that this component can be responsible for
modulating the infalling plasma flow near the star's surface with
FLRs' frequencies and producing the observed X-ray fluxes, see Eq.
(\ref{par-comp}) below. We will return to this point later.

\begin{figure}[h]
\begin{center}
\includegraphics[angle=0,width=8cm]{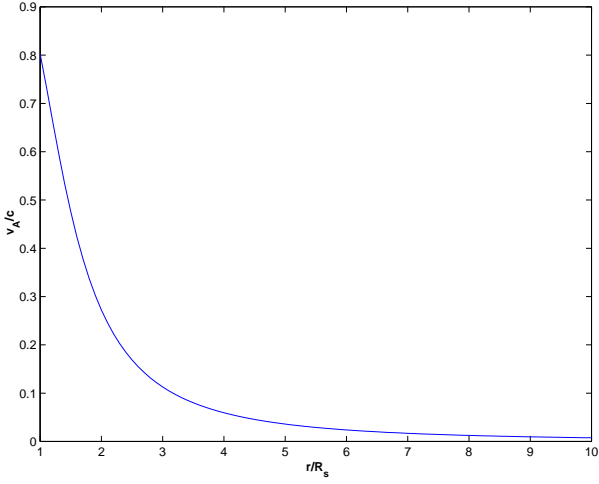}
\caption{Radial profile of the relativistic Alfv\'en speed $v_{\rm A}$
(in units of speed of light) in the equatorial plane, see Eq. (\ref{rel_Alf_vel}).
$R_s$ is the radius of
the neutron star.  Here we assumed $p=0$, $R_s=10$ km, $M=M_\odot$,
$\dot{M}=10^{17}$ g s$^{-1}$ and $\mu=10^{26}$ G cm$^3$.}
\label{dip_va2e}
\end{center}
\end{figure}

In a strong magnetic field, the usual definition of the Alfv\'en
velocity, $v_{\rm A}^2=B^2/(4\pi\rho)$, is not correct as with
this non-relativistic definition, velocities can be
larger than the speed of light, $c$. The correct definition,
the relativistic Alfv\'en velocity, uses energy density
$\epsilon$ relative to $c^2$, which for an ideal gas reads
\st
\begin{equation}\label{enthalpy}
h\equiv \frac{\epsilon}{c^2} = \frac{c^2 \rho + \gamma p/(\gamma-1)+B^2/(4\pi)}{c^2}
\end{equation}
rather than the plasma density $\rho$. Consequently the (relativistic)
Alfv\'en velocity is
\st
\begin{equation}\label{rev_alfv}
v_{\rm A}^2=B^2/(4\pi h)= \frac{c^2 }{c^2 \rho + \gamma p/(\gamma-1) +
B^2/(4\pi)}~\frac{B^2}{4\pi}\,.
\end{equation}
In Fig. \ref{dip_va2e}, we plot radial profiles of the relativistic
Alfv\'en velocity $v_{\rm A}$ in the equatorial plane (assuming $p=0$).
As shown in this figure, close to the star the relativistic
Alfv\'en velocity approaches the speed of light, ie. $v_{\rm A}
\sim c$.  This definition for the Alfv\'en velocity is used extensively
in studies of relativistic jet dynamics in black hole systems and quasars
(see for an example \citealt*{Miz04}).
Again, we use a linear approximation
to obtain a dispersion relation with resonant coupling.

The linearized perturbed magnetohydrodynamic
equations in the presence of an ambient flow can be obtained from Eqs. (\ref{mhd_eq}) as
\st\begin{eqnarray}\label{eq-mhd1}
\stq\label{xi}
&& h \lp {\partial \delta \v \over  \partial t}
+  \v \cdot \nab
\delta\v +  \delta\v \cdot \nab \v\rp =
 - \nab \delta P   + {\nab P\over h}\delta h \no\\
&&\hspace{1cm}+{1\over 4\pi}\lp\frac{}{}\delta \B \cdot \nab \B +\B
\cdot \nab \delta \B\frac{}{}\rp, \\
\stq\label{dB_t}
&&{\partial \delta \B\over\partial t} = \nab \times (\delta \v \times \B + \v
\times \delta \B), \\
\stq\label{div_dB}
&&\nab\cdot\delta \B =0, \\
\stq\label{eqs-2}
&&({\partial \over \partial t} + \v\cdot\nab)\delta p + \delta \v \cdot \nab p = - \gamma
( \delta p \nab \cdot \v + p  \nab \cdot \delta \v),\no\\
\end{eqnarray}
where $\delta \v=\partial \xxi/\partial t$.    $P=p+B^2/(8\pi)$
and $\delta P=\delta p+\B\cdot\delta\B/(4\pi)$ are the total unperturbed and perturbed
plasma pressure (fluid + field), respectively.
Here $h, p, \v,$ and $\B$ are the unperturbed quantities
while $\delta h, \delta p, \delta\v,$ and $\delta\B$ are the perturbed quantities.
Furthermore, we consider the slow rotation approximation and
neglect both the toroidal field $B_\phi$ and velocity $v_\phi$,
Eq. (\ref{comp}), to avoid complexities.   Such assumptions may not meet the actual
configuration precisely.  Nevertheless, these assumptions simplify
our calculations significantly.

To analyze the problem analytically, we consider again a
rectilinear magnetic field configuration.  We note that although
this configuration may not be suitable for the accreting neutron
star, it provides us with a descriptive picture that can be applicable
to QPOs.
Separating Eqs. (\ref{eq-mhd1}) into parallel and perpendicular
components relative to the ambient magnetic field and assuming
perturbed quantities in the form of
\st
\begin{equation}
\delta (\r,t) = \delta (\r_\perp) e^{-i (\omega t - k_{||} r_{||})}\,,
\end{equation}
we find
\st \bea\label{eq-mhd2}
\stq\label{par-comp}
&&-i\omega_D\xi_{||} =
-\frac{k_{||}v^2_{\rm s}}{\omega_D}\nab\cdot\xxi -
(\xxi_\perp\cdot\nab_\perp)v_{||} \,,\\
\stq\label{perp-comp}
&&(\omega_D^2-k_{||}^2v_{\rm
A}^2) \xxi_\perp  =
\frac{\omega_D}{\omega}\frac{\nab_\perp\delta P}{h}
-\frac{\omega_D}{\omega}\frac{\nab_\perp P}{h^2}\delta h
\,,\\
\stq\label{dB_par}
&&-i \omega_D\delta B_{||}= i\omega
B(\nab_\perp\cdot \xxi_\perp) + i\omega
(\xxi_\perp\cdot\nab_\perp) B\no\\
&&~~~~~~~+(\delta\B_\perp\cdot\nab_\perp) v_{||}\,,~~{\rm (Using \nab\cdot\delta\B=0)}\\
\stq\label{dB_perp}
&&-i \omega_D \delta\B_\perp = \omega k_{||} B \xxi_\perp\,,
~~~~\hspace{0cm}{\rm (Using \nab\cdot\B=0)} \\
\stq\label{dp}
&&-i \omega_D \delta p=i\omega (\xxi_\perp\cdot\nab_\perp)p
-\gamma \omega p( k_{||}\xi_{||} -i
\nab_\perp\cdot\xxi_\perp),\\
\stq\label{PP}
&&-\frac{\delta
P}{h}=\frac{\omega}{\omega_D}(v_{\rm s}^2+v_{\rm
A}^2)\nab\cdot\xxi -\frac{i\omega k_{||} v_{\rm
A}^2}{\omega_D}\xi_{||} \no\\
&&\hspace{3.5cm}+\frac{\omega k_{||} v_{\rm
A}^2}{\omega_D^2}
(\xxi_\perp\cdot\nab_\perp)v_{||}\,,
\eea
where $\omega_D=\omega-k_{||}v_{||}$ is the Doppler shifted frequency, $\omega$ is the
eigenfrequency, $k_{||}$ is the wave number in the parallel
direction, $\v_p=v_{||}\B/B$, $v_{\rm s}=\sqrt{\gamma p/h}$ is the
sound velocity \footnote{In order to obtain Eq. (\ref{PP}) we
assumed that at equilibrium $\nab P=\nab(p+B^2/(8\pi))=0$. This is
valid for a non-rotating or very slowly rotating star.}.  Note that we assumed
that all ambient quantities are function of $\r_\perp$ only.
In order to derive Eq. (\ref{eq-mhd2}),  Eq. (\ref{par-comp})
uses Eqs. (\ref{dB_perp}) and (\ref{dp}),
Eq. (\ref{perp-comp}) uses
Eq. (\ref{dB_perp}), and Eq. (\ref{PP}) is a combination of Eqs.
(\ref{dB_par}), (\ref{dB_perp}), and (\ref{dp}).
Equation (\ref{par-comp}) shows that the plasma displacement
parallel to the ambient field does not vanish even in the cold
plasma approximation ($p=0$)\footnote{In case of zero ambient
flow, $v_{||}=0$, the parallel displacement $\xi_{||}$ vanishes in
the cold plasma}. The non-zero $\xi_{||}$ can affect and then
modulate the motion of plasma along the field lines. This
modulation will occur at the frequency of the resonant shear Alfv\'en
waves $\omega$ and should be observed in X-ray fluxes.

We set the magnetic field in the z-direction with
gradients in the ambient parameters in the x-direction ($\r_\perp$) only, i.e.
$v_{||}(\r_\perp)=v_p(x)=v_z(x)$, and further
\st
\begin{equation}
\delta (\r,t) = \delta(x) e^{-i (\omega t - k_y y - k_z z)}\,.
\end{equation}
Here we assumed that the perturbed quantities have sinusoidal dependencies
in both y- and z-directions.
Equations (\ref{par-comp}), (\ref{perp-comp}) and (\ref{PP}) in
the cold plasma approximation ($p=0$) can then
be combined into one differential equation
\st\begin{eqnarray}
\label{line}
&& \frac{d^2
\xi_x}{dx^2} + \frac{\omega_D\Xi '(x)+k_zv'_p\Xi (x)}{\omega_D\Xi (x)}
\frac{d \xi_x}{dx}\no\\
&& +
\lp\frac{k_zv'_p\Xi'(x)}{\omega_D\Xi (x)}
+\lp \frac{k_zv'_p}{\omega_D}\rp' 
+ \frac{\omega_D^2 - (k_y^2+k_z^2)v_{\rm A}^2}{v_{\rm A}^2}\rp
\xi_x = 0,\no\\
\end{eqnarray}
where $\omega_D=\omega-k_zv_p$ and
\st\be\label{Xi}
F(x)\equiv \omega_D\Xi(x) =
\frac{\omega_D^2 - k_z^2 v_{\rm A}^2}
{\omega_D^2 - (k_y^2 + k_z^2)  v_{\rm A}^2 } ~h~ v_{\rm A}^2\,.
\ee
Note that here $v_p=v_p(x)$ and $v_{\rm A}=v_{\rm A}(x)$ are
functions of $x$.  For $v_p=0$ Eqs. (\ref{line}) and (\ref{Xi}) reduce to the one
obtained in section 2.2.1, Eqs. (\ref{har}), (\ref{G1}) and (\ref{F1})
with $\rho_0$ replaced by $h$, as expected.

We note that mathematically our result is similar to one obtained by \cite{HS92},
however, they considered a different model in which
flow in the Earth's magnetosheath is perpendicular to the B-field and there
is no ambient flow in the region of FLRs.
For a constant flow along magnetic field lines, i.e. $v'_p(x)=0$,
Eq. (\ref{line}) reduces to
\st\begin{eqnarray}
\label{line1}
\frac{d^2
\xi_x}{dx^2} + \frac{F'(x)}{F (x)}\frac{d \xi_x}{dx}
+ \frac{\omega_D^2 - (k_y^2+k_z^2)v_{\rm A}^2}{v_{\rm A}^2}
\xi_x = 0\,.
\end{eqnarray}
Equation (\ref{line1}) has a singular point at $F=0$.
The condition $F=0$ yields the following modes:
\st
\be\label{om_A}
\omega^{\pm}=  k_z (v_p \pm v_{\rm A}) .
\ee
The resulting modes are combination of a
compressional mode $k_zv_p$ due to the plasma flow and a shear
Alfv\'en mode $k_zv_{\rm A}$.   They produce two frequency peaks
in the spectrum that vary with time if the velocities change with
time. The value of $k_z \sim \pi/L$ where $L$ is the length of
the field line and is on the order of the radius of the neutron star, $L
\sim
R_s$.  Therefore, for $R_s\sim 10^6$ cm and $v_{\rm A}\sim v_p\sim
1.5\times 10^8$ cm s$^{-1}$ we get $\omega^+ \sim 1000$ Hz, which is
comparable to the frequencies of the observed QPOs.
In the limit $v_p\ll v_{\rm A}$ these modes become a
single mode of a regular Alfv\'en resonance. In the case of
$v_{\rm A}\ll c$  this mode becomes consistent with a single mode
(\ref{flres}).   For a superalfv\'enic flow, ie. $ v_{\rm A}\ll
v_p$, however, the MHD Alfv\'en waves and the resulting FLRs are
suppressed due to propagation of the MHD wave,
$k_zv_p$.  In the case of superalfv\'enic plasma flow, one has
\st\be
F(x) \simeq
\frac{h k_z^2 v_{\rm A}^2}{k_y^2+k_z^2},~~ {\rm as }~~
\omega\rightarrow k_z v_p\,,
\ee
which shows that Eq. (\ref{line1}) is no longer singular.
Therefore, no FLRs are likely to occur in this case.

As a result, we expect that
the two FLR peaks (Eq. \ref{om_A}) occur where $v_p$ and $v_{\rm A}$ are
more or less comparable (i.e. at the same order of magnitude).  As we now show,
using the flow velocity and the Alfv\'en velocity definitions below,
one might expect that the FLRs likely occur at $R_s < r <
100 R_s$.

Approximating the plasma inflow velocity with the free fall
velocity $v_p\sim v_{\rm ff}(r)$ and $v_{\rm A}\sim
B(r)/\sqrt{4\pi \rho_{\rm ff}}$ where $\rho_{\rm
ff}=\dot{M}/(v_{\rm ff}~ 4\pi r^2)$ is the free fall mass density,
one can rewrite Eq. (\ref{om_A}) as:
 \st \bea \omega^{\pm}(r)&\simeq&  k_z
\lp\frac{}{}
(2GM/r)^{1/2} \pm B(r)/\sqrt{4\pi \rho_{\rm ff}(r)}\frac{}{}\rp,\no\\
&\simeq& k_z(v_p(R_s) x^{-1/2} \pm v_{\rm A}(R_s) x^{-9/4})\,,
\eea
where $v_p(R_s)=(2GM/R_s)^{1/2}\simeq 1.6\times 10^{10}
(M/M_\odot)^{1/2}(R_s/10\,{\rm km})^{-1/2}$ cm s$^{-1}$ and
$v_{\rm A}(R_s)=B(R_s)/\sqrt{4\pi\rho_{\rm ff}(R_s)}\simeq 4 \times
10^{10} \mu_{26} \dot{M}_{17}^{-1/2} (M/M_\odot)^{1/4}
(R_s/10\,{\rm km})^{-9/4}$ cm s$^{-1}$ are inflow and Alfv\'en
velocities at the surface of the star\footnote{We note that to
calculate the Alfv\'en velocity at the surface of the star
properly, one  would need to use the relativistic Alfv\'en velocity as
defined in Eq. (\ref{rev_alfv}), i.e.
\st
\be\label{rel_Alf_vel}
v_{\rm A}(r)=\frac{\mu
\dot{M}^{-1/2}(2GM)^{1/4}r^{-9/4}}{\sqrt{1+\mu^2
\dot{M}^{-1}(2GM)^{1/2}r^{-9/2}/c^2}}.\ee
Here we assumed $p=0$.  Using the above
relation the Alfv\'en velocity at the surface of the star will be
$\sim 0.8 c$, $.997 c$, and $c$ for $\mu=10^{26},\,10^{27}$, and
$10^{28}$ G cm$^3$, respectively.  At further distances from the
star, however, the relativistic Alfv\'en velocity reduces to its
classical version  which we used instead.
The values given here for the relativistic
Alfv\'en velocity at the stellar surface, are for a neutron star with
$R_s=10$ km, $\dot{M}=10^{17}$
g s$^{-1}$ and $M=M_\odot$}. Here $\mu_{26}$ is the magnetic field dipole
moment at the surface of the star in units of $10^{26}$ G cm$^{3}$,
$\dot{M}_{17}$ is the mass of accretion rate in units of $10^{17}$
g s$^{-1}$, $x=r/R_s$, and $R_s$ is the radius of the star. Figures
\ref{freq_vs_r} and \ref{freq_vs_alfv} show  the regular
Alfv\'en resonance frequencies ($v_p=0$) or FLRs, as a
function of distance from the star and the Alfv\'en
speed. It is clear that the closer to the star and/or the
larger the Alfv\'en velocity the higher the frequency. Therefore,
for an Alfv\'en speed say $v_{\rm A}(R_s)\sim .1 c$ one can get the
frequency about $1000$ Hz at $r\simeq 2.78 R_s$ as we expected.
Furthermore, since the Alfv\'en velocity depends on the energy density
(or density) of the plasma which varies from time to time due to
several processes such as magneto-turbulence at boundaries during
the accretion, the resulting FLRs frequencies will also vary with
time.  In addition, the position where the resonance takes place,
is also subject to change in time due to the time varying
accretion rate.   Such time varying behavior causes time varying
frequencies or quasi-periodic frequencies as observed.

Setting $\omega^-$ as the observed peak separation frequency
$\Delta \nu\sim \omega^-$, and $\omega^+$ as the upper QPO
frequency $\nu_2\sim\omega^+$,
one can find $\nu_1,~\nu_2$ and $\Delta\nu$ as
\st\bea
\stq\label{nu_1}
&&\nu_1(x)\simeq 5k_zv_p(R_s)\mu_{26}\dot{M}_{17}^{-1/2}
(M/ M_\odot)^{-1/4}\no\\
&&\hspace{3.5cm}\times \lp R_s/ 10~{\rm km}\rp^{-7/4}x^{-9/4},\\
\stq\label{nu_2}
&&\nu_2(x)\simeq k_zv_p(R_s)[x^{-1/2}+\no\\
&&\hspace{0cm}2.5\mu_{26}\dot{M}_{17}^{-1/2}
\lp M/ M_\odot\rp^{-1/4}\lp R_s/ 10~{\rm km}\rp^{-7/4}x^{-9/4}],\\
\stq\label{d_nu}
&&\Delta\nu(x)\simeq k_zv_p(R_s)[x^{-1/2}+\no\\
&&\hspace{0cm}2.5\mu_{26}\dot{M}_{17}^{-1/2}
\lp M/ M_\odot\rp^{-1/4}\lp R_s/ 10~{\rm km}\rp^{-7/4}x^{-9/4}].
\eea
Figures \ref{peak_freq_low} and \ref{peak_freq}
show the variation of $\Delta\nu$ vs $\nu_1$ and $\nu_2$,
respectively, along with the observed values.   The curves are drawn
by assuming $\mu_{26}=.32$ with (top to bottom)
$k_zv_p(R_s)\sim 860$
Hz, $750$ Hz, and $650$ Hz, respectively. As shown by
observations, the value of $\Delta\nu$ decreases whenever the
magnitude of $\nu_2$ decreases (sufficiently) and/or it increases.
Such behavior is expected in our model (see Fig. \ref{peak_freq}).

The lower QPO frequency can also be rewritten as
$\nu_1=\nu_2 - \Delta\nu\sim 2k_z v_{\rm A}$.   As a result,
we obtain the frequency ratio $\nu_1$ to $\nu_2$
\st
\be\label{om_ratio} \nu_1/\nu_2= 2v_{\rm A}/(v_p + v_{\rm A})=
2/(1+v_p/v_{\rm A}) \,.
\ee
It is clear that depending on the
relation between the plasma flow velocity $v_p$ and the Alfv\'en
velocity $v_{\rm A}$, one can obtain different frequency ratios.
For example, if the velocity of the inflow plasma is $2$ ($7/3\sim 2.3$)
times greater than the Alfv\'en velocity, i.e.
$v_p=2v_{\rm A}~(v_p=7/3~v_{\rm A})$, one
can get $2:3$ ($3:5$) ratio. For a free falling plasma along the
lines of a dipolar magnetic field, $v_p(r)\propto r^{-1/2}$,
whereas $B(r)/\sqrt{4\pi\rho(r)}\propto r^{-9/4}$, the condition
$v_p\sim 2v_{\rm A}$ ($v_p\sim 2.3v_{\rm A}$) is satisfied at
$r\sim 3 R_s\sim 1.5 r_{\rm A}$ (here $\mu_{26}=1$ is assumed).
The occurrence of FLRs is very
likely to be at this distance from the star.

Such pairs of high frequencies with a $2:3$ frequency ratio have
been observed in the X-ray flux of some neutron stars in
LMXBs\footnote{In Sco X-1 the correlation line between two
frequencies has a steeper slope than $2/3$ {\bf (see \citealt*{Bel04})}.} and black hole
systems (the $3:5$ ratio has seen in one source), and it would
appear that his feature is common to those systems.
{\bf However, recent statistical analyzes by \citet{Bel04,Bel05} showed that
although the two high and low kHz QPO frequencies in LMXBs are well correlated,
the frequency-frequency correlation is significantly different from
a $2:3$ relation.  They analyzed all published values of kHz QPO frequencies
from neutron star systems and found that the strong linear correlation observed
between the lower and upper kHz QPO frequencies in both Atoll and Z sources
is not compatible with a single constant ratio.
Furthermore, by analyses of a much larger RXTE/PCA dataset for Sco X-1, they showed that
there is no sharp concentration around a $2:3$ ratio, but that the ratios
are broadly distributed over the range $820-1150$ Hz.
}

The
existence of such rational ratios is still a mystery. In black
hole systems, it has been suggested that such frequencies
correspond to a trapped g-mode or c-mode of disk oscillation in
the Kerr metric, see for example \cite{Kat01} and references
therein.  In neutron star systems they are explained as the
fundamental and the first harmonic of the non-axisymmetric ($m=1$)
g-mode \citep{Kat02,Kat03}.
Further, \citet{Rez03a,Rez03b} studied small
perturbations of an accretion
torus orbiting close to the black hole and modeled the observed
high QPO frequencies with basic p-modes of relativistic tori.
They showed that these modes behave as sound waves trapped in
the torus with eigenfrequencies appearing in sequence 2:3:4:...
\cite{Abr03} also proposed that
the observed rational ratios of frequencies may be due to the
strong gravity of the compact object and a non-linear resonance
between radial and vertical oscillations in accretion disks.


\section{Discussion}\label{conc}

In the present work, we have studied the interaction of an
accretion disk with a neutron star magnetosphere in the LMXBs. The
recent extensive observations reveal the existence of
quasi-periodic oscillations in the X-ray fluxes of such stars.
These oscillations, with frequencies ranging from $10$ Hz to $1200$
Hz, have been the subject of several theoretical and observational
investigations. Based on theoretical models for the observed
aurora in the Earth's magnetosphere, we have introduced a generic
magnetospheric model for accretion disk-neutron star systems to
address the occurrence and the behavior of the observed QPOs in
those systems. In order to explain those QPOs consistently, we
consider the interaction of the accreting plasma with the neutron
star magnetosphere.   Due to the strong gravity of the star, a
very steep and supersonic flow hits the magnetosphere boundary and
deforms its structure drastically. Such a plasma flow can readily
excite different MHD waves in the magnetosphere, including shear
Alfv\'en waves.

\begin{figure}[h]
\begin{center}
\includegraphics[angle=0,width=8cm]{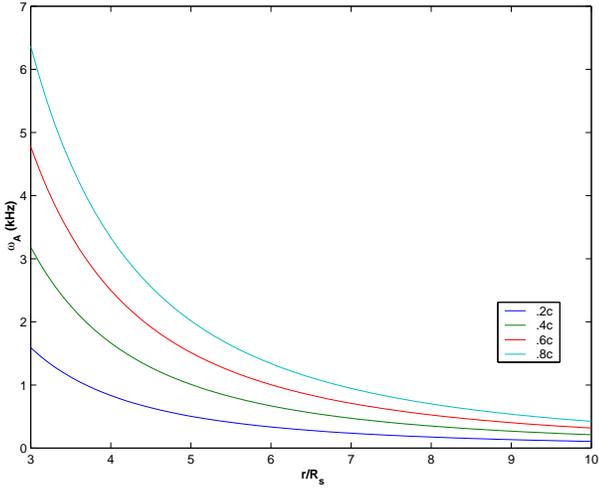}
\caption{Frequency of the regular Alfv\'en resonance
($v_p=0$) as function of $r$,
$\omega_{\rm A} =k_z v_{\rm A}(R_s) (r/R_s)^{-9/4}$. The closer to
the star the higher the frequency. From bottom to top
$v_{\rm A}(R_s)=.2c,~ .4c,~ .6c$ and $.8c$, respectively. Here $R_s=10$ km,
and $k_z=\pi/R_s$.}
\label{freq_vs_r}
\end{center}
\end{figure}

\begin{figure}[h]
\begin{center}
\includegraphics[angle=0,width=8cm]{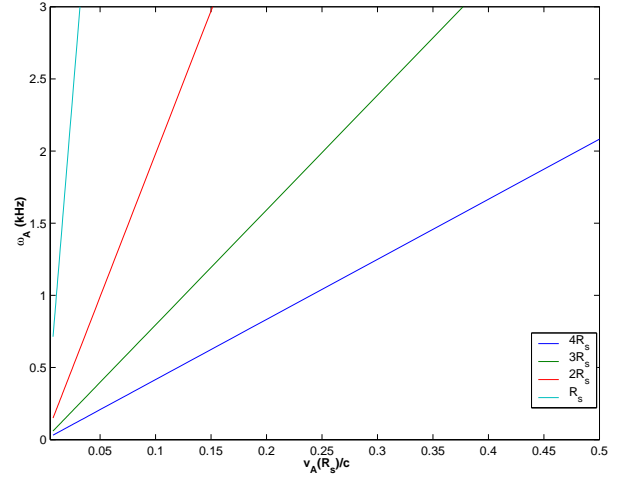}
\caption{Frequency of the regular Alfv\'en resonance ($v_p=0$) as a function of
the Alfv\'en wave speed at the stellar radius, $\omega_{\rm A} =k_z v_{\rm A}(R_s) (r/R_s)^{-9/4}$.
The larger the Alfv\'en speed the
higher the frequency. From top to bottom
$r/R_s=1, 2, 3$, and $4$, respectively. Here $R_s=10$ km,
and $k_z=\pi/R_s$.  } \label{freq_vs_alfv}
\end{center}
\end{figure}

\begin{figure}[h]
\begin{center}
\includegraphics[angle=0,width=8cm]{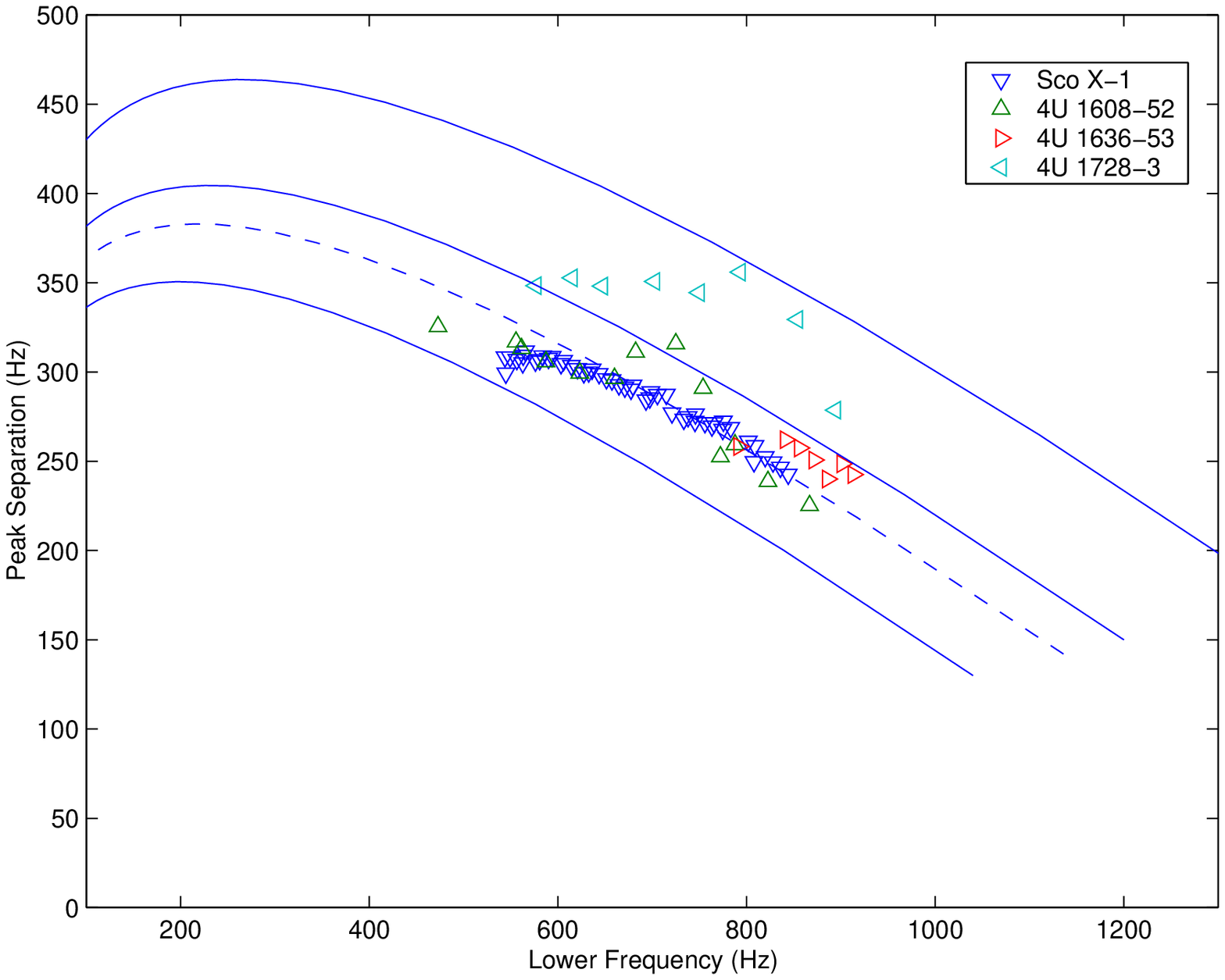}
\caption{Variation of the spectral peak separation
$\Delta\nu$ with
the lower frequency $\nu_1$, see Eqs. (\ref{nu_1}) and (\ref{d_nu}).
The resulting curves are compared
with the observed values for different systems.
The solid curves
from bottom to top correspond to $k_zv_p(R_s)\sim 650,\,750,$ and
$860$ Hz. The dashed curve corresponds to $k_zv_p(R_s)\sim 710$
Hz.  Here we assumed $\mu_{26}=.32,~ R_s=10$ km, $\dot{M}_{17}=1$,
and $M=M_\odot$.} \label{peak_freq_low}
\end{center}
\end{figure}

\begin{figure}[h]
\begin{center}
\includegraphics[angle=0,width=8cm]{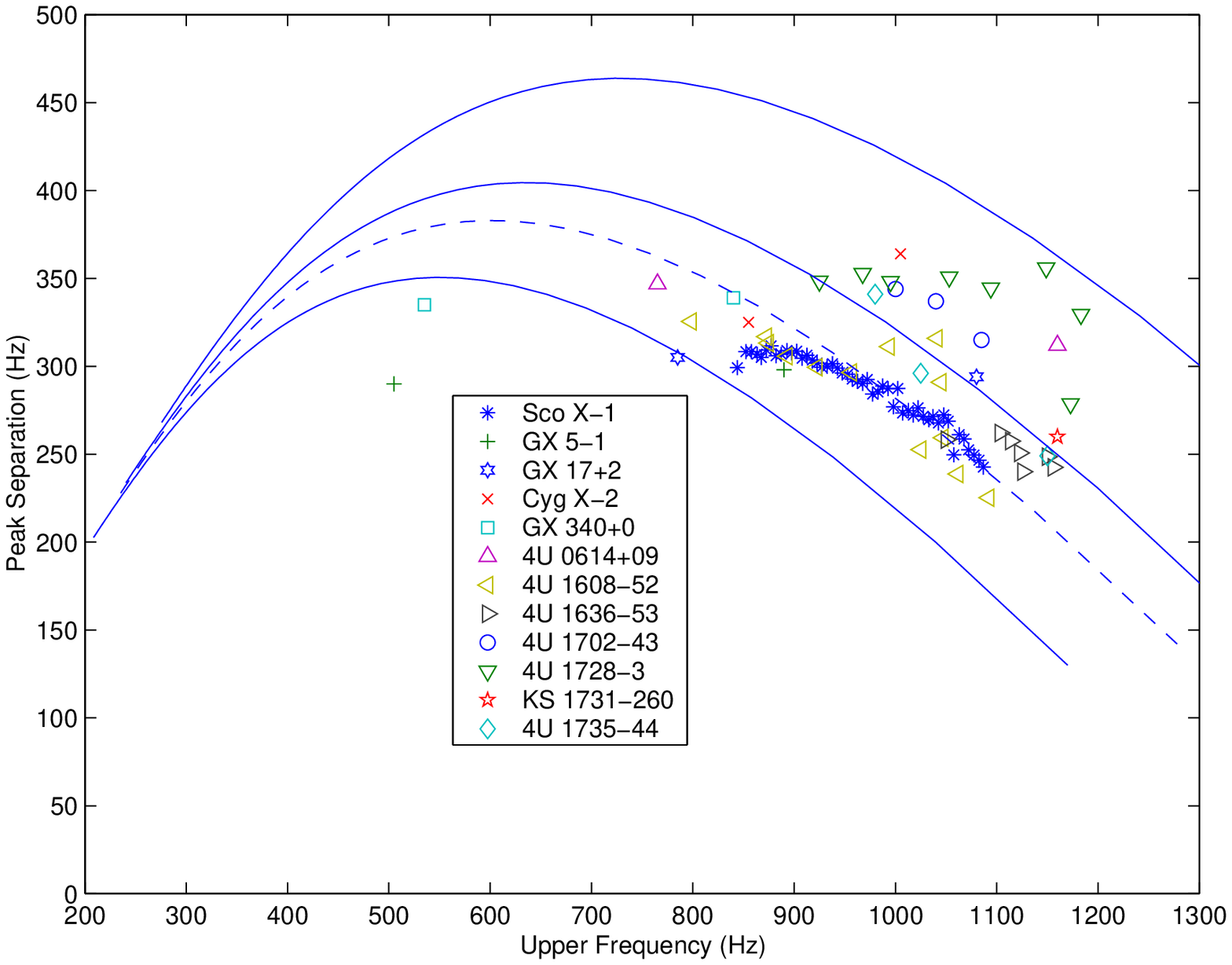}
\caption{ Variation of the spectral peak separation
$\Delta\nu$ with the upper frequency $\nu_2$,
see Eqs. (\ref{nu_2}) and (\ref{d_nu}).
The resulting curves are compared
with the observed values for different systems.
The solid curves
from bottom to top correspond to $k_zv_p(R_s)\sim 650,\,750,$ and
$860$ Hz. The dashed curve corresponds to $k_zv_p(R_s)\sim 710$
Hz. Here we assumed $\mu_{26}=.32,~ R_s=10$ km, $\dot{M}_{17}=1$,
and $M=M_\odot$. } \label{peak_freq}
\end{center}
\end{figure}

In the Earth's magnetosphere, occurrence of aurora is a result of
the resonant coupling between the shear Alfv\'en waves and the fast
compressional waves (produced by the solar wind). These resonances
are known as FLRs and are reviewed in detail in section
\ref{sect:mag} of this paper. We argue that such resonant coupling
is likely to occur in neutron star magnetospheres due to the
interaction with accreting plasma.  In the context
of QPOs in neutron star magnetospheres, we formulated an improved FLR
model by considering a plasma flow moving with velocity $v_p$ along the
magnetic field lines. Such flows are likely to occur in a neutron star
magnetosphere \citep{GLP77}. For a simple geometry, a
rectilinear magnetic field, and in the presence of a plasma flow
we found: (a) two resonant MHD modes with
frequencies $\omega^{\pm} = k_z(v_p \pm v_{\rm A})$. (b) the
resulting frequencies for $k_z\sim \pi/R_s\sim 3\times 10^{-6}$
cm$^{-1}$ and typical flow and/or Alfv\'en velocity $\sim .1 c$
at the stellar surface will be in kHz range within a few stellar radii.
Our results match the kHz oscillations
observed in the X-ray fluxes in LMXBs. As shown in figures
\ref{freq_vs_r} and \ref{freq_vs_alfv}, the closer to the star
and/or the larger the Alfv\'en velocity the higher the frequency.
(c) the quasi-periodicity of the observed oscillations can be
understood by noting that due to several processes such as
magneto-turbulence at boundaries and the time varying accretion
rate, the FLR frequencies may vary with time. (d) a non-zero
plasma displacement along the magnetic field lines $\xi_{||}$.
Such a displacement, which oscillates with resulting frequencies
$\omega^\pm$, modulates the flow of the plasma toward the surface
of the neutron star.  As a result, the X-ray flux from the star will
show these frequencies as well. (e) setting $\Delta\nu=\omega^-$
and $\nu_2=\omega^+$, one can explain the behavior of the
peak separation frequency $\Delta\nu$ relative to the upper QPO
frequency $\nu_2$.  As observed, the value of $\Delta\nu$
decreases as the magnitude of $\nu_2$ decreases and/or increases.
Figure \ref{peak_freq} clearly shows such behavior. (f) for
$v_p\sim 2 v_{\rm A}$ (obtained at $r\sim 3 R_s$ for $\mu_{26}\sim 1$),
the frequency ratio
$\nu_1/\nu_2$ is comparable with the observed frequency ratio
$2:3$ in some neutron star systems.  {\bf However, a broad range of
frequency ratios (\citealt*{Bel04,Bel05}) is expected as values
of $v_p$ and $v_{\rm A}$ change.
}

Interestingly, using the observed values of QPO frequencies, one
can determine the mass density and the magnetic dipole moment
of the star using
\st
\bea
\stq
&&k_z v_p(r)=(1/2)(\nu_2+\Delta\nu)=\nu_2-\nu_1/2,\\
\stq
&&k_z v_{\rm A}(r)=(1/2)(\nu_2-\Delta\nu)=\nu_1/2\,.
\eea
Therefore, \st \bea \label{den}
\stq
&&M_s/R_s^3 \simeq \frac{1}{8\pi^2 G} x^3 (\nu_2+\Delta\nu)^2 ,\\
\stq &&\mu_s\simeq  (32\pi^4 G)^{-1/4} R_s^{5/2} \dot{M}^{1/2}
(M_s/R_s^3)^{-1/4} x^{13/4} \nu_1, \eea where $k_z\simeq \pi/r$,
$\mu_s=B_0R_s^3$,$x=r/R_s$, and $B_0$ is the magnetic field
strength at the surface of the star.   In table 1, we calculate
the average density $M_s/R_s^3$ and magnetic dipole moment $\mu_s$
of the star for a fixed value $x=10$. Note that larger frequencies
may occur at smaller distances, ie. $x<10$.  Our results are
compatible with realistic neutron star parameters.

Furthermore, our model is able to explain the low frequency
($\sim 10$ Hz) quasi-periodic oscillations observed in the rapid
burster such as MXB 1730-335 and GRO J1744-28 \citep{Mas00}.  Such
low frequencies can be extracted from the model by considering
smaller inflow/Alfv\'en velocities and/or further distances from the
star.

Nevertheless, in order to avoid a number of complexities in our
calculations, we used approximations such as slow rotation and
a cold plasma.  These assumptions may put some restriction on the
validity of our model and results.  Future studies will be devoted
to overcoming those restrictions.
\acknowledgements
VR wishes to thank Mariano Mendez for kindly providing QPO data.
The authors appreciate the referee's careful reading
of the manuscript and valuable suggestions.
This research was supported by the National Sciences and
Engineering Research Council of Canada.

%
%
\def\aj{{AJ}}                   
\def\araa{{ARA\&A\ }}             
\def\apj{{ApJ\ }}                 
\def\apjl{{ApJ\ }}                
\def\apjs{{ApJS\ }}               
\def\apss{{Ap\&SS}}             
\def\aap{{A\&A\ }}                
\def\aapr{{A\&A~Rev.}}          
\def\aaps{{A\&AS}}              
\def\azh{{AZh}}                 
\def\baas{{BAAS}}               
\def\jrasc{{JRASC}}             
\def\memras{{MmRAS}}            
\def\mnras{{MNRAS\ }}             
\def\pra{{Phys.~Rev.~A}}        
\def\prb{{Phys.~Rev.~B}}        
\def\prc{{Phys.~Rev.~C\ }}        
\def\prd{{Phys.~Rev.~D\ }}        
\def\pre{{Phys.~Rev.~E}}        
\def\prl{{Phys.~Rev.~Lett.\ }}    
\def\pasp{{PASP}}               
\def\pasj{{PASJ\ }}               
\def\qjras{{QJRAS}}             
\def\skytel{{S\&T}}             
\def\solphys{{Sol.~Phys.}}      
\def\sovast{{Soviet~Ast.\ }}      
\def\ssr{{Space~Sci.~Rev.\ }}     
\def\zap{{ZAp}}                 
\def\nat{{Nature\ }}              
\def\iaucirc{{IAU~Circ. No.}}       
\def\aplett{{Astrophys.~Lett.}} 
\def\apspr{{Astrophys.~Space~Phys.~Res.}}
\def\bain{{Bull.~Astron.~Inst.~Netherlands}}
\def\fcp{{Fund.~Cosmic~Phys.}}  
\def\gca{{Geochim.~Cosmochim.~Acta}}   
\def\grl{{Geophys.~Res.~Lett.}} 
\def\jcp{{J.~Chem.~Phys.}}      
\def\jgr{{J.~Geophys.~Res.}}    
\def\jqsrt{{J.~Quant.~Spec.~Radiat.~Transf.}}
\def\memsai{{Mem.~Soc.~Astron.~Italiana}}
\def\nphysa{{Nucl.~Phys.~A}}   
\def\nphysb{{Nucl.~Phys.~B\ }}   
\def\physrep{{Phys.~Rep.}}   
\def\physscr{{Phys.~Scr}}   
\def\planss{{Planet.~Space~Sci.}}   
\def\procspie{{Proc.~SPIE}}   

----------------------------------------------------------------------------------------

\newpage

%
%
\begin{table}[htbp]
\caption{Observed frequencies of kilohertz QPOs in Z and atoll
sources} \label{table1} \scriptsize
\begin{center}
\begin{tabular}{lccclcc}
\hline\hline
         &       &$\nu_1$ &$\nu_2$& $\Delta \nu$ & $M_s/R_s^3$ & $\mu_s$\\
Source   & Type  &(Hz)    & (Hz) & (Hz)& ($10^{14}$ g cm$^{-3}$)& ($10^{26}$ G cm$^3$)\\
\hline
Sco\,X$-$1     & Z  & 565 & 870  & 307$\pm$5  &  2.6  & 6.6  \\
GX\,5$-$1      & Z  & 660 & 890  & 298$\pm$11 &  2.7 & 7.6\\
GX\,17+2       & Z  & 780 & 1080 & 294$\pm$8  &  3.6  & 8.4\\
Cyg\,X$-$2     & Z  & 660 & 1005 & 346$\pm$29 &  3.5  & 7.2 \\
GX\,340+0      & Z  & 565 & 840  & 339$\pm$8  &  2.6  & 6.6 \\
GX\,349+2      & Z  & 710 & 980  & 266$\pm$13 &  2.9   & 8.0 \\
4U\,0614+09    & atoll  & 825    & 1160  & 312$\pm$2 &  4.1 & 8.6\\
4U\,1608$-$52  & atoll  & 865    & 1090  & 225$\pm$12 & 3.3 & 9.5\\
4U\,1636$-$53  & atoll  & 950    & 1190  & 251$\pm$4  & 3.9 & 10.0\\
4U\,1702$-$43  & atoll  & 770    & 1085  & 315$\pm$11 & 3.7 & 8.2\\
4U\,1705$-$44  & atoll  & 775    & 1075  & 298$\pm$11 & 3.6 & 8.3\\
4U\,1728$-$34   & atoll  & 875    & 1160  & 349$\pm$2  & 4.3 & 9.0\\
KS\,1731$-$260 & atoll  & 900    & 1160  & 260$\pm$10 & 3.8 & 9.5\\
4U\,1735$-$44  & atoll  & 900    & 1150  & 249$\pm$15 & 3.7 & 9.6\\
4U\,1820$-$30  & atoll  & 795    & 1075  & 278$\pm$11 & 3.5 & 8.6\\
Aql\,X$-$1     & atoll  & 930    & 1040  & 241$\pm$9  & 3.1 & 10.3\\
4U\,1915$-$05  & atoll  & 655    & 1005  & 348$\pm$11 & 3.5 & 7.2\\
XTE\,J2123$-$058& atoll & 845   & 1100   & 255$\pm$14 & 3.5 & 9.2\\
\hline
\end{tabular}
\end{center}

\vbox{ Low ($\nu_1$) and high ($\nu_2$) QPO frequencies with
corresponding peak separation ($\Delta\nu$) observed in Z and
atoll sources. Only one data series is given as example, for
further detail see \cite{Van00}. Values of $M_s/R_s^3$ and $\mu_s$
are calculated from Eq. (\ref{den}) for a fixed value $x=10$ as an example.
Here we assumed ${\dot M}=10^{17}$ g s$^{-1}$ and $R_s=10$ km.
We note that larger frequencies may occur at smaller distances,
$x<10$. }
\end{table}


\clearpage

\end{document}